\documentclass[12pt]{article}
\usepackage[all]{xy}

\usepackage{amsmath,amssymb,amscd,bm,mathtools,array}

\usepackage[margin=1.75cm]{caption} 
\usepackage[labelfont=it,textfont={small,it}]{caption} 

\usepackage{hyperref}
\usepackage{cite}

\let\ssection=\section
\renewcommand{\section}{\setcounter{equation}{0}\ssection}

\newcommand{\diag}{\mathrm{diag}}
\newcommand{\cO}{{\mathcal{O}}}

\newcommand{\barP}{\overline{P}}
\newcommand{\bp}{{\mathbf{p}}}
\newcommand{\np}{\Vert\bp\Vert}
\newcommand{\bq}{{\mathbf{q}}}
\newcommand{\Pf}{\mathrm{Pf}}

\newcommand{\bs}{{\mathbf{s}}}
\newcommand{\ns}{\Vert\bs\Vert}

\newcommand{\Tr}{\mathrm{Tr}}

\newcommand{\dX}{\dot{X}}
\newcommand{\bx}{{\mathbf{x}}}
\newcommand{\bsigma}{{\boldsymbol{\sigma}}}
\newcommand{\half}{\frac{1}{2}}

\title{
How does the photon's spin affect Gravitational Wave measurements?}

\author{
Lo\"ic Marsot\footnote{mailto: loic.marsot@cpt.univ-mrs.fr}\\[8pt]
Aix Marseille Univ, Universit\'e de Toulon, CNRS, CPT,
Marseille, France
}

\begin{document}
\maketitle

\begin{abstract}
We study the effect of the polarization of light beams on the time delay measured in Gravitational Wave experiments. To this end, we consider the Mathisson-Papapetrou-Dixon equations in a gravitational wave background, with two of the possible spin supplementary conditions: by Frenkel-Pirani, or by Tulczyjew. In the first case, photons follow a null geodesic and thus no spin effect is present. The second case shows a deviation of the photons from the null geodesic, resulting in a tiny effect on the measured time delay of photons depending on their polarization state.
\end{abstract}

\section{Introduction}
Gravitational wave detection in interferometers such as the Laser Interferometer Gravitational-Wave Observatory (LIGO) and the Virgo observatory involves laser beams traveling through a gravitational field perturbed by a gravitational wave inhomogeneity. The wave profile is reconstructed from the difference of time of flight of the laser light in two perpendicular linear arms. Theoretically, the time of flight is computed by treating the beam as a collection of photons, with each photon moving on a geodesic in a given (gravitational wave) background. In general, however, geodesics are only followed by {\em spinless} particles. In the present paper, we thus try to include the photons' spin and check whether it could lead to a measurable effect.

In general relativity, the motion of spinning test particles is described by the Mathisson-Papapetrou-Dixon (MPD) equations \cite{Mat37,Pap51,Dix70,Tul59}. See \cite{Sou74} for a completely geometrical description of these equations. These follow from the treatment of extended test bodies based on the multipole expansion along a certain worldline, \textit{a posteriori} identified as representing the history of the body. Such an expansion has sense when the length scales connected with the body are much shorter than the curvature length scale. If all the multipoles beyond dipole are neglected, the system reduces to the MPD equations. One then speaks of a ``pole-dipole'' approximation. The MPD equations are given by,

\begin{align}
\dot{P}^\mu & = - \half {R^\mu}_{\rho\alpha\beta} S^{\alpha \beta} \dot{X}^\rho, \label{mpdp} \\
\dot{S}^{\mu\nu} & = P^\mu \dot{X}^\nu - P^\nu \dot{X}^\mu, \label{mpds}
\end{align}
where $X$, $P$ and $S$ denote respectively the position, momentum and spin
tensor of the test particle. The dot over the trajectory $X$ denotes the ordinary derivative with respect to its affine parameter, $\dot{X} = dX/d\tau$, while the dot over $P$ and $S$ denotes the covariant derivative with respect to that same parameter.

Note that the MPD equations do not determine the evolution uniquely: we lack an equation for $\dot{X}$ (the latter needs not be parallel to $P$). This
reflects an ambiguity in the selection of the worldline $X(\tau)$ representing
the particle history. One thus has to impose certain constraints to close the system. These can be written in the form ${S^\mu}_\nu V^\nu=0$, where $V^\mu$ is a suitable vector. These constraints are usually called spin supplementary conditions (SSCs). The vector $V^\mu$ may in principle be chosen freely, though there are several obvious ``intrinsic'' options, provided by the geometry of the problem itself. In the present paper, we consider two of such possibilities, the Mathisson-Pirani (or Frenkel-Pirani) SSC: $V \parallel \dot{X}$ \cite{Mat37,Pir56,Fre26}, and the Tulczyjew SSC: $V \parallel P$ \cite{Tul59,Dix70}. While the former is known to generally keep the spinning photon on a geodesic, we show that the latter leads to nontrivial spin effects.

The lack of constitutive laws which determine how the body responses to gravitational and inertial strains leads to the freedom which the different supplementary conditions fix, each in a different way. In particular, they lead to different trajectories. It is not possible yet to say which of the conditions is the ``correct'' one. See \cite{KyrianS07,KunstLLGS15,HarmsLGBN16,LGHBN17,WitzanySLG18} for comparison of different SSCs.

In the past, the Mathisson-Pirani SSC was sometimes deemed unsatisfactory due to there not being a unique representative worldline, depending on the choice of initial conditions \cite{Mat372,Wey47,Moller49}. This issue has been clarified recently in \cite{CostaNZ12,CostaHNZ11,CostaLGS17}, in connection with discovering the momentum-velocity relation for that SSC. The Tulczyjew SSC, on the other hand, does provide a unique worldline, irrespectively of how the initial conditions are prescribed \cite{Dix70}.

Since $\dot{X}$ needs not be parallel to $P$ anymore, the theory naturally offers different definitions of the body's ``mass'', $m=\sqrt{P^\mu P_\mu}$, $\tilde{m}=\dot{X}^\mu P_\mu$, and possibly $V^\mu P_\mu$. The MPD equations by themselves do not ensure that any of the above masses stays constant, not even that the vectors $P$ and $\dot{X}$ are, or remain, timelike. However, we obtain more information with the help of the chosen SSC. For the Mathisson-Pirani SSC ${S^\mu}_\nu\dot{X}^\nu=0$, it is $\tilde{m}$ that stays conserved. Setting this to zero for a massless particle, implies that the particle follows a null geodesic, while the momentum vector is spacelike \cite{Mas75,BailynR81,BiniCAJ06,Semerak15}. For the Tulczyjew SSC ${S^\mu}_\nu P^\nu=0$, it is $m$ that is conserved. Setting this to zero does not necessarily imply a null geodesic for the photon, the more so that the vector $\dot{X}$ may in this case become space-like \cite{Sat76}.

Let us add that the MPD equations ensure, independently of the SSC, the conservation, along the representative worldline, of the spin-tensor invariant $2s^2=S_{\alpha\beta}S^{\alpha\beta}$. This scalar is sometimes called the longitudinal spin and for photons it equals $\pm\hbar$, with sign$(s)$ called helicity or handedness. By fixing the conserved mass and spin, this completes the description of a classical elementary particle as belonging to one of the coadjoint representations of the Poincar\'e group.

In the massless case, the choice of the SSC is even more subtle than in the massive one. Two main arguments have been given in favor of the Mathisson-Pirani SSC: i) Maxwell equations minimally coupled to gravity yield null geodesics in the geometric-optic limit \cite{Laue20}, like do the MPD equations together with this SSC \cite{Mas75,BailynR81,BiniCAJ06} (with just one type of counterexample given in \cite{BailynR81}). (ii) Imposing conformal invariance of the theory, in particular the tracelessness of the energy-momentum tensor, implies (a slight generalization of) the Mathisson-Pirani constraint \cite{Duv78,BailynR77,Semerak15}. Less satisfactorily, the MPD equations supplemented with that constraint do not behave well in the $\tilde{m}\rightarrow 0$ limit, the massless problem is actually unrelated to the massive one \cite{Wey47,Obu11}. On the other hand, Tulczyjew’s SSC has often been considered inappropriate because, as already mentioned, it generally leads to a spacelike motion, which is more serious than the spacelike momentum yielded by the Mathisson-Pirani SSC. It also leads to a certain degeneracy of the massless problem in flat spacetime: rather than a localized particle, it yields a plane traveling at the speed of light.

Recently, however, the Tulczyjew SSC has been revisited in connection with phenomena observed in spinoptics. As already predicted by Fedorov and Imbert \cite{Fed55,Imb72}, the wave packet of spinning light should perform an ``instantaneous'' transverse shift when being reflected at an interface. This effect can be described theoretically using the symplectic mechanics in a 3-dimensional manifold \cite{Duv06,Duv07,Duv08,Duv13} similar to the symplectic representation of Souriau’s spinning-particle model involving the Tulczyjew SSC \cite{Sou74}. The effect, also called Spin Hall Effect of light, was confirmed experimentally in 2008 \cite{Hos08,Bli08}. Recall that Fermat’s principle can be rephrased like that the light rays follow null geodesics in a 3-dimensional Riemannian space conformally related to the Euclidean one by a scale factor represented by the local refractive index squared. One can then summarize the 2008's experiments as follows: the spinning light rays deviate from null geodesics in the above space. More specifically, the speed of spinning light can locally become higher than the speed of spinless light, without violating causality over distances larger than the wavelength of the photon.

Also in favor of the Tulczyjew SSC, one can mention the presence of the Berry phase in quantum mechanics, which is in general connected with a deviation from geodesics as well. In specific examples, the treatment of the problem with the help of a Berry phase and the treatment with the MPD equations with the Tulczyjew SSC, or their symplectic description, agree with each other. See, for instance \cite{StoneDZ14,DuvalH14} for the treatment of chiral fermions, and \cite{Gos06,ChDLMTS} for birefringence of a photon in a Schwarzschild spacetime (note that there is a typo in the very last formula in \cite{Gos06}: their anomalous velocity is indeed transverse to the geodesic plane, just as in \cite{ChDLMTS}). Still another support for the Tulczyjew SSC was provided by Souriau who showed \cite{Sou70} that geometric quantization of the symplectic system which derives the MPD equations with this SSC, when considered with a flat background, leads to the Maxwell equations.

To summarize, the MPD equations with the Tulczyjew SSC may provide an effective, semi-classical description of phenomena tied to the photon spin and involving the occurrence of faster speeds than that of spinless light. Note that if causality is not violated over distances larger than the wavelength of the photon, it should not imply any problem, since the pole-dipole approximation as such only holds if the length scales tied to the particle (here wavelength of the photon) are much smaller than the curvature length scale. Indeed, in papers where the Tulczyjew SSC was employed, e.g., to study photons in the Schwarzschild, de Sitter or FLRW backgrounds \cite{Sat76,ChDTSRW,ArmazaHKZ16,ChDLMTS}, causality has not been found to be violated over meaningful distances.

Note that the Tulczyjew SSC has already been used in the problem of massive spinning-particle motion in an exact gravitational wave solution \cite{Moh01}. In \cite{ObuSilAleTer17}, classical as well as quantum massive fermions were studied, with application to a gravitational-wave background (among others).

Let us add, finally, that gravitational birefringence had already been considered experimentally in 1974 \cite{Har74}, resulting in an upper bound for this effect in gravitational lensing, but the results were somewhat inconclusive, since the effect can actually be expected to be much weaker than the experiment precision \cite{Gos06,ChDLMTS}. Thanks to the high sensitivity of LIGO and Virgo, experimental bounds can also be found for birefringence predicted by other theories, for example those violating the Lorentz invariance \cite{KosMelMew16,KosMew16}.

The paper is organized as follows. In section \ref{s:notations}, we introduce the notation. In section \ref{s:UR}, the main point is explained which is to consider a photon in a gravitational-wave background as a limit of an ultrarelativistic particle traveling in one direction. Section \ref{s:eom} contains the computations needed to obtain the equations of motion in a weak-field approximation. And we conclude by summary and comments in section \ref{s:conclusions}.

\section{Notations}
\label{s:notations}
First, let us introduce our notations. The metric has signature $(-, -, -, +)$. The components of the Riemann curvature tensor are defined by the convention ${R^\mu}_{\nu\alpha\beta} = \partial_\alpha \Gamma^\mu_{\beta\nu} - \partial_\beta \Gamma^\mu_{\alpha\nu} + \cdots$. In this paper, we often suppress indices by considering linear maps instead of 2-tensors. For instance, we use the linear map $S = ({S^\mu}_\nu)$ and likewise for the shorthand notation $R(S)$, with ${R(S)^\mu}_\nu = {R^\mu}_{\nu\alpha\beta} S^{\alpha\beta}$. In the same way, we have the vector $P$ and the associated covector $\barP = (\barP_\mu)$ where indices are lowered with the metric. Another shorthand notation will be $R(S)(S)=R_{\mu\nu\alpha\beta}S^{\mu\nu}S^{\alpha\beta}$. 

For a skew-symmetric linear map $F$, the operator $\Pf$ gives its Pfaffian $\Pf(F)$. With the fully skew-symmetric Levi-Civita tensor $\epsilon_{\mu\nu\rho\sigma}$, with $\epsilon_{1234} = 1$, we have the expression $\Pf(F) = - \frac{1}{8} \sqrt{-\det(g_{\alpha\beta})} \epsilon_{\mu\nu\rho\sigma} F^{\mu\nu}F^{\rho\sigma}$. We have the relation $\Pf(F)^2 = \det(F)$. Indeed, the determinant of a skew-symmetric matrix can always be written as a perfect square.

\section{Photons as a limit of ultrarelativistic particles}
\label{s:UR}

The so-called Souriau-Saturnini equations are the combination of the MPD equations, together with the Tulczyjew constraint $SP=0$, and applied to the case of the photon. For massless particles, we consider the momentum such that $P^2=0$, and so, for $R(S)(S) \neq 0$, we have the equations (Souriau\cite{Sou74} and Saturnini\cite{Sat76} in French, see \cite{ChDTSRW} for the proof in English),
\begin{align}
\dot{X} & = P+\frac{2}{R(S)(S)}S R(S) P\,, \label{xdot_massless}\\
\dot{P} & = -s\,\frac{\Pf(R(S))}{R(S)(S)}\,P\,,\\
\dot{S} & = P\overline{\dX}-\dX\barP. 
\end{align}

The Souriau-Saturnini equations describe the trajectory of a massless photon with spin in a gravitational field. While they work rather well in a Robertson-Walker background \cite{ChDTSRW}, or in the proximity of a star \cite{Sat76,ChDLMTS}, they break down when the curvature of the gravitational background vanishes. This is due to the lonely term $R(S)(S)$ in the denominator of (\ref{xdot_massless}). When the curvature vanishes, the equations become those of a plane wave traveling at the speed of light. Indeed, massless and chargeless particles cannot be localized in flat spacetime with this approach. It becomes a problem for a metric of gravitational waves, as they are usually computed as a perturbation around flat spacetime.

This time, for massive particles, $P^2 = m^2 \neq 0$, and we have similar equations \cite{Kun72,Obu11},
\begin{align}
\dot{X} & = P - \,\frac{2 \, S R(S) P}{4 \, P^2 - R(S)(S)}, \label{xdot_massive}\\
\dot{P} & = - \half R(S) \dot{X},\label{pdot}\\
\dot{S} & = P\overline{\dX}-\dX\barP.\label{sdot}
\end{align}

Notice that we recover the Souriau-Saturnini equations in the limit $P^2 \rightarrow 0$, which is not, \textit{a priori}, trivial. For example, this is not the case with the Pirani constraint.

Now, for massive particles, the denominator of (\ref{xdot_massive}) behaves in a nicer way. When the Riemann tensor goes to zero, or when $m^2 \gg R(S)(S)$, we recover the usual geodesic equation. To be sure the denominator does not vanish in the massive case, we should have $4 m^2 > R(S)(S)$. We thus have a lower bound on the mass of the test particle. With $f$ the frequency of the gravitational wave and $c$ the speed of light, that requirement becomes
\begin{equation}
m^2 > \frac{\epsilon \,\pi^2 \, f^2 \, \hbar^2}{c^4}
\end{equation}
Note that this depends on the amplitude $\epsilon$ of the gravitational waves. As this amplitude goes to zero, the mass restriction reduces to $m > 0$. In the case of gravitational wave detections, the frequency of gravitational waves is typically around $f = 50$Hz, and the amplitude around $\epsilon = 10^{-20}$. This gives
\begin{equation}
\label{m_constraint}
m > 10^{-59}\,\mathrm{ kg},
\end{equation}
to have a consistent set of equations describing a massive particle with spin in a typical background with gravitational waves.

The main idea to compute the time delay due to the photon's spin in a background of gravitational waves is to only compute the effect in the direction defined by the momentum. Indeed, the photon goes back and forth in one direction of propagation, so here we are not interested in the full trajectory in space of the photon/particle.  Therefore, to compute the delay, we can compute the effect of spin on a massive particle, though with a mass much smaller than its momentum. Since we only compute the time delay in the direction defined by the momentum, and since (\ref{xdot_massive}) reduces to (\ref{xdot_massless}) in the limit $P^2 \rightarrow 0$, the mass will drop out of the equations when compared to the momentum, thus giving us the expected time of flight delay for a photon.

Notice that, in any case, the best experimental measurements on the mass of a photon give us an upper bound for the mass of about $10^{-50}\,$kg to $10^{-54}\,$kg depending on the type of measurements and assumptions \cite{Wil71,Ryu07}. These upper bounds are a few orders of magnitude higher than the constraint on the mass of the photon (\ref{m_constraint}) in the massive equations.

\section{Equations of motion for the ultrarelativistic photon}
\label{s:eom}
Using Cartesian coordinates $(x_1,x_2,x_3,t)$, we linearize the gravitational field equations with the metric, 
\begin{equation}
\label{metric_gw}
g_{\mu\nu}= \eta_{\mu\nu} + \epsilon \, h_{\mu\nu} + \cO(\epsilon^2)
\end{equation}
where $(\eta_{\mu\nu}) = \diag(-1,-1,-1,1)$ is the flat Minkowski metric, $h_{\mu\nu}$ the linear deviation of the metric to flat spacetime, and $\epsilon \ll 1$ a small parameter encoding the amplitude of the gravitational wave.

Linearizing the Einstein field equations in $\epsilon$, and considering a gravitational wave propagating in the direction of the $z$ axis, leads to the well-known solution for the perturbation~$h_{\mu\nu}$,
\begin{equation}
(h_{\mu\nu}) = 
\left(
\begin{array}{cccc}
f_+(t-x_3) & f_\times(t-x_3) & 0 & 0 \\
f_\times(t-x_3) & -f_+(t-x_3) & 0 & 0 \\
0 & 0 & 0 & 0 \\
0 & 0 & 0 & 0
\end{array}
\right)
\end{equation}
with $f_+$ and $f_\times$ two functions describing the two polarization states of the gravitational waves.

For concreteness, take $f_+(t-x_3) = \cos(\omega(t-x_3))$ and $f_\times(t-x_3) = 0$ with $c = 1$. The linearized metric thus takes the form,
\begin{equation}
\label{g_e}
(g_{\mu\nu}) = 
\left(
\begin{array}{cccc}
-1 + \epsilon \cos(\omega(t-x_3)) & 0 & 0 & 0 \\
0 & -1 - \epsilon \cos(\omega(t-x_3)) & 0 & 0 \\
0 & 0 & -1 & 0 \\
0 & 0 & 0 & 1
\end{array}
\right)+ \cO(\epsilon^2)
\end{equation}

Up to linear order in $\epsilon$, we have ${R^3}_{1 3 1} = -{R^3}_{1 4 1} = - {R^3}_{2 3 2} = {R^3}_{2 4 2} = {R^4}_{1 3 1} = - {R^4}_{1 4 1} = - {R^4}_{2 3 2} = {R^4}_{2 4 2} = - \half \omega^2 \epsilon \cos(\omega(t-x_3))$. 

Now, to alleviate notations, we write $k \equiv \cos(\omega(t-x_3))$. The conditions $P^2 = m^2$, and to recover the usual four-momentum $P$ in the limit $\epsilon \rightarrow 0$, dictate the expression,
\begin{equation}
(P^\mu) = \left(
\begin{array}{c}
\displaystyle p_1 \left(1 + \frac{\epsilon}{2} k\right) \\[1.3ex]
\displaystyle p_2 \left(1 - \frac{\epsilon}{2} k\right) \\[1.3ex]
\displaystyle p_3 \\[1.3ex]
\sqrt{m^2 + \np^2}
\end{array} 
\right)+ \cO(\epsilon^2),
\end{equation}
where the $p_i = p_i(t), i = 1, 2, 3$ are the unknown components of the 3-momentum, and with $\np^2 = p_1^2 + p_2^2 + p_3^2$.

Likewise, the spin tensor is defined by its constraints. To linear order in $\epsilon$ we have, with $s_i = s_i(t)$ understood,
\begin{equation}
\label{def_s}
({S^\mu}_\nu)=
\left(
\begin{array}{cccc}
0 & -s_3\left(1 + \epsilon k\right) & s_2\left(1 + \frac{\epsilon}{2} k\right) & \frac{(p_2 s_3 - p_3 s_2)}{\sqrt{m^2+\np^2}}\left(1 + \frac{\epsilon}{2} k\right) \\
s_3\left(1 - \epsilon k\right) & 0 & -s_1\left(1 - \frac{\epsilon}{2} k\right) & \frac{(p_3 s_1 - p_2 s_3)}{\sqrt{m^2+\np^2}} \left(1 - \frac{\epsilon}{2} k\right) \\
-s_2\left(1 - \frac{\epsilon}{2} k\right) & s_1\left(1 + \frac{\epsilon}{2} k\right) & 0 & \frac{(p_1 s_2 - p_2 s_1)}{\sqrt{m^2+\np^2}} \\
\frac{(p_2 s_3 - p_3 s_2)}{\sqrt{m^2+\np^2}}\left(1 - \frac{\epsilon}{2} k\right) & \frac{(p_3 s_1 - p_2 s_3)}{\sqrt{m^2+\np^2}}\left(1 + \frac{\epsilon}{2} k\right) & \frac{(p_1 s_2 - p_2 s_1)}{\sqrt{m^2+\np^2}} & 0 
\end{array}
\right)
\end{equation}
such that $S$ is skew-symmetric, and still up to linear order,
\begin{equation}
SP = 0 \qquad \mathrm{and} \qquad -\half \Tr(S^2) = j^2
\end{equation}
with
\begin{equation}
j^2 = \frac{(\bs\cdot\bp)^2+m^2 \ns^2}{\np^2+m^2}.
\end{equation}
Note that in the limit $m \rightarrow 0$ in the above relation, we recover the square of the scalar spin, or longitudinal spin, of a massless particle. In other words, in the massless case, the longitudinal spin is the projection of the spin vector along the direction of the momentum.

Next, we have,
\begin{equation}
\Pf(R(S)) = \cO(\epsilon^2).
\end{equation}

See Appendix \ref{AppendixA} for the expressions of $R(S)(S)$ and $S\,R(S)\,P$.

We then have the equations of motion for the position of the massive particle (\ref{xdot_massive}),
\begin{equation}
\dot{X} = P - \,\frac{2 \, S R(S) P}{4 \, P^2 - R(S)(S)},
\end{equation}

So, we get the equations of motion on 3d-space, with respect to the time coordinate $t$, in the 3+1 splitting $(\bx,t)$, as
\begin{equation}
\frac{d \bf x}{dt} = \frac{\left(2 m^2-\half R(S)(S)\right) {\bf P}-{\bf SR(S)P}}{\left(2 m^2-\half R(S)(S)\right) P_4-SR(S)P_4}
\end{equation}

At this point, the mass terms allow us to take the limit $\epsilon \rightarrow 0$.  Indeed, as we will see below. From \eqref{pdot} and \eqref{sdot}, which we can rewrite as equations for $d\bp/dt$ and $d\bs/dt$ with the $3+1$ split, we see that $d\bp/dt \sim d\bs/dt \sim \cO(\epsilon)$. Hence, if we take the following initial conditions for the photon,
\begin{equation}
\bx_0 = \left(\begin{array}{c}
0 \\
0 \\
0
\end{array}
\right), \quad \bp_0 = \left(\begin{array}{c}
0 \\
{p_2}_0 \\
0
\end{array}
\right), \quad \bs_0 = \left(\begin{array}{c}
{s_1}_0 \\
{s_2}_0 \\
{s_3}_0
\end{array}\right),
\end{equation}
we have the following momentum and spin, $\bp(t) = \bp_0 + \epsilon \, \bq(t) + \cO(\epsilon^2)$ and $\bs(t) = \bs_0 + \epsilon \, \bsigma(t) + \cO(\epsilon^2)$. Since we only want the equation of motion $dx_2/dt$ at linear order in $\epsilon$, it is sufficient to have $\bp(t)$ and $\bs(t)$ at the zeroth order in $\epsilon$. Indeed, as we will see below, contributions in $\bq(t)$ and $\bsigma(t)$ vanish after the ultrarelativistic limit.

Thus, for the velocity in the direction we are interested in, at first order in $\epsilon$, we have,
\begin{align}
\frac{dx_2}{dt} = & \frac{{p_2}_0}{\sqrt{m^2+{p_2}_0^2}} + \epsilon \frac{m^2 q_2(t)}{(m^2+{p_2}_0^2)^{3/2}} + \nonumber \\
& - \frac{\epsilon}{2} \, \frac{{p_2}_0(m^2+{p_2}_0^2) + \omega^2 \Big({p_2}_0({s_1}_0^2-{s_3}_0^2) - \sqrt{m^2+{p_2}_0^2} {s_2}_0 {s_3}_0 \Big)}{(m^2+{p_2}_0^2)^{3/2}} \, \cos(\omega(t-x_3)) + \cO(\epsilon^2)
\end{align}

We might be interested here in the behaviour of the function $q_2(t)$. From \eqref{pdot} and the $3+1$ split, we get, 
\begin{equation}
\frac{d q_2(t)}{dt} = \frac{\epsilon}{2} \omega \big({p_2}_0 \sin(\omega(t-x_3)) - {s_1}_0 \omega \cos(\omega(t-x_3))\big).
\end{equation}
The important take here is that $q_2(t)$ does not contain any mass term. Thus, when ${p_2}_0^2 \gg m^2$, we have
\begin{equation}
\label{dxdt_ur}
\frac{d x_2}{dt} = 1 - \frac{\epsilon}{2} \cos(\omega(t-x_3)) - \frac{\epsilon}{2} \frac{\lambda_\gamma^2}{\lambda_{\mathrm{GW}}^2} \frac{\left({s_1}_0^2-{s_3}_0^2-{s_2}_0{s_3}_0\right)}{\hbar^2} \cos(\omega(t-x_3)) + \cO(\epsilon^2)
\end{equation}
with $\lambda_\gamma$ the wavelength associated to the photon, and $\lambda_{\mathrm{GW}} = 2\pi/\omega$ is the wavelength of the gravitational wave. With values taken from LIGO/Virgo, $\lambda_\gamma = 1064$nm, 
\begin{equation*}
\frac{\epsilon}{2} \frac{\lambda_\gamma^2}{\lambda_{\mathrm{GW}}^2} \sim 10^{-46}.
\end{equation*}

This means that geodesic effects of order $\epsilon^2 \sim 10^{-40}$ would be seen before observing any spin effect in LIGO/Virgo type detectors.

The effect is maximum when photons are polarized such that $\bs = (0, \hbar, \hbar)$, at least in the classical limit. In that case, the measured time delay is decreased from~$\Delta\tau$ to
\begin{equation}
\widetilde{\Delta\tau} = \Delta\tau\left(1 - 2 \frac{\lambda_\gamma^2}{\lambda_{\mathrm{GW}}^2}\right)
\end{equation}

A corollary is that two photons of different polarization will have different times of flight. Thus, a beam made up of photons of random polarization will introduce a noise due to spin curvature effects. A way to eliminate this noise is to polarize the beams of light before sending them into the arms. However, the amplitude of the noise created by this birefringence is of the relative order of $10^{-46}$ in LIGO/Virgo, which is much below the current sensitivity in LIGO and Virgo experiments.

\section{Conclusions}
\label{s:conclusions}

To take into consideration the possible effects of the photon's spin on its trajectory in curved space, we used the Mathisson-Papapetrou-Dixon equations for spinning test particles, together with two possible supplementary conditions for photons, by Frenkel-Pirani, or by Tulczyjew. While for a massive spinning body, such as a spinning star, the choice of SSC does not seem to have much practical impact on the observable trajectory (unless the angular momentum of the body is extremely large \cite{Semerak99}), the choice for elementary particles has more consequences. 

The Frenkel-Pirani SSC for a massless particle leads to a trajectory along a null geodesic, regardless of the gravitational background. In that case, there would be no change to the geodesic trajectory of photons in a background of gravitational waves.

The Tulczyjew SSC for a massless particle predicts a very small effect due to the polarization of the light on its trajectory. Since the massive equations with this condition lead to the massless equations in the limit $m \rightarrow 0$, and because of the instability of the localization of the test particle in the equations near zero curvature, the photon is treated in this paper as an ultrarelativistic massive particle. This mass, which can be both large compared to the spin-curvature coupling term $R(S)(S)$ and extremely small compared to the momentum of the photon, allows for convenient limits to be taken in the equations. The geodesic equations in a gravitational wave background are recovered, together with a new term depending on the spin polarization of the photon. This means that with this supplementary condition, the time of flight of a photon in a detector depends on its polarization state. This dependence is, however, many order of magnitudes lower than the first order effects of gravitational waves on the time of flight. But, if we achieve that kind of precision, polarizing the laser beam in a specific way would be an easy way to reduce the noise introduced by birefringence. With enough precision, this could even potentially be a way to discriminate between the two possible Spin Supplementary Conditions.

\vskip 12ex
\paragraph{Acknowledgements:} The project leading to this publication has received funding from Excellence Initiative of Aix-Marseille University - A*MIDEX, a French ``Investissements d'Avenir'' programme.

\vbadness=10000

\renewcommand\thesection{A}
\numberwithin{equation}{section}

\section{Appendix: Expressions of R(S)(S) and SR(S)P}\label{AppendixA}
From the expression of the Riemann tensor, of the spin tensor (\ref{def_s}), $k \equiv \cos(\omega(t-x_3))$, and $R(S)(S)~=~R_{\mu\nu\lambda\sigma}S^{\mu\nu}S^{\lambda\sigma}$, we get,

\begin{equation}
\begin{split}
R(S)(S) = \frac{2 \omega^2 \epsilon k}{m^2+\np^2}& \Big[
2 (p_1 s_1 - p_2 s_2) s_3 \left(p_3 - \sqrt{m^2+\np^2}\right) - \left(p_1^2-p_2^2\right)s_3^2 + \\
& - \left(s_1^2 - s_2^2\right) \left(p_3 \left(p_3 - 2 \sqrt{m^2+\np^2}\right)+\left(m^2+\np^2\right)\right)
\Big]+ \cO(\epsilon^2).
\end{split}
\end{equation}

Similarly, we obtain, with $SR(S)P^\mu = {R^\mu}_{\nu\lambda\sigma} P^\nu S^{\lambda\sigma}$,

\begin{equation}
SR(S)P = \left(\begin{array}{c}
SR(S)P_1 \\
SR(S)P_2 \\
SR(S)P_3 \\
SR(S)P_4
\end{array}\right),
\end{equation}
with,

\begin{align}
SR(S)P_1 = & K \Bigg(s_3 \left(m^2+\np^2\right) \left(\sqrt{m^2+\np^2}-p_3\right) \left(s_1 \left(\sqrt{m^2+p^2}-p_3\right)+p_1 s_3\right) + \nonumber \\
& -\left(s_2 \sqrt{m^2+\np^2} \left(\sqrt{m^2+\np^2}-p_3 \right)+p_2 s_3 \right) \times \\
& \times \left(\left(\sqrt{m^2+\np^2}-p_3\right) \left(p_2 s_1+p_1 s_2\right)+2 p_1 p_2 s_3\right)\Bigg) + \cO(\epsilon^2), \nonumber \\
SR(S)P_2 = & K \Bigg(s_3 \left(m^2+\np^2\right) \left(p_3-\sqrt{m^2+\np^2}\right) \left(s_2 \left(\sqrt{m^2+\np^2}-p_3\right)+p_2 s_3\right) + \nonumber \\
& +\left(s_1 \sqrt{m^2+\np^2} \left(\sqrt{m^2+\np^2}-p_3 \right)+p_1 s_3 \right) \times \\
& \times \left(\left(\sqrt{m^2+\np^2}-p_3\right) \left(p_2 s_1+p_1 s_2\right)+2 p_1 p_2 s_3\right)\Bigg) + \cO(\epsilon^2), \nonumber \\
SR(S)P_3 = & K \sqrt{m^2+\np^2} \Bigg(\left(s_2^2-s_1^2\right) \sqrt{m^2+\np^2} \left(\sqrt{m^2+\np^2}-p_3\right)^2 +\nonumber \\
& + \left(\sqrt{m^2+\np^2}-p_3\right) \left(-s_3 \sqrt{m^2+vp^2} \left(p_1 s_1-p_2 s_2\right)+p_2^2 s_1^2-p_1^2 s_2^2\right)+\\
& + 2 p_1 p_2 s_3 \left(p_2 s_1-p_1 s_2\right) \Bigg) + \cO(\epsilon^2), \nonumber \\
SR(S)P_4 = & K \sqrt{m^2+\np^2} \Bigg(2 p_3^2 \left(\sqrt{m^2+\np^2}-p_3\right) \left(s_1^2-s_2^2\right)+s_3 \left(p_1^3 s_1-p_2^3 s_2\right) + \nonumber \\
& +3 s_3 p_1 p_2 \left(p_2 s_1-p_1   s_2\right)+\left(m^2+3 p_3^2-3 \sqrt{m^2+\np^2} p_3\right) s_3 \left(p_1 s_1-p_2 s_2\right) + \\
& -\left(\sqrt{m^2+\np^2}-2 p_3\right) \left(p_1^2 s_2^2-p_2^2   s_1^2\right)+\left(\sqrt{m^2+\np^2}-p_3\right) s_3^2 \left(p_1^2-p_2^2\right) + \nonumber \\
& -p_3 \left(p_1^2 s_1^2-p_2^2 s_2^2\right)- m^2 p_3 \left(s_1^2-s_2^2\right)\Bigg) + \cO(\epsilon^2), \nonumber 
\end{align}
and,
\begin{equation}
K = \frac{\omega^2 \epsilon \cos (\omega (t-x_3)) }{\left(m^2+\np^2\right)^{3/2}}.
\end{equation}

\end{document}